\documentclass[12pt]{spie} 

\pdfoutput=1

\usepackage{epsfig}
\usepackage{verbatim}
\usepackage{color}
\usepackage{amsmath}
\usepackage{amssymb}

\usepackage{longtable}
\usepackage{braket}
 \usepackage{caption}
 \usepackage{subcaption}
 \usepackage{multirow}
 \usepackage{longtable,lscape}
 \usepackage{color}

 \author{V. Lutsker\supit{a}, B. Aradi\supit{b}, and Thomas A.\
   Niehaus\supit{a}\footnote{$\;$ Corresponding author:
    thomas.niehaus@ur.de}
 \skiplinehalf
 \supit{a} Department of Theoretical Physics, University of
   Regensburg, 93040 Regensburg, Germany\\
 \supit{b} BCCMS, University of Bremen, 28359 Bremen, Germany}
\title{Implementation and benchmark of a long-range corrected functional
in the density functional based tight-binding method}

\begin{document}
\maketitle

\begin{abstract}
Bridging the gap between first principles
methods and empirical schemes, the density functional based tight-binding method (DFTB) has become a versatile tool in predictive
atomistic simulations over the past years. One of the major
restrictions of this method is the limitation to local or gradient
corrected exchange-correlation functionals. This excludes the important class of
hybrid or long-range corrected functionals, which are advantageous in
thermochemistry, as well as in the computation of vibrational,
photoelectron and optical spectra. The present work provides a detailed account of the implementation of DFTB
for a long-range corrected functional in generalized Kohn-Sham theory. We apply the method
to a set of organic molecules and compare ionization potentials and
electron affinities with the original DFTB method and higher level
theory. The new scheme cures the significant overpolarization in electric
fields found for local DFTB, which parallels the functional dependence
in first principles density functional theory (DFT). 
At the same time the computational savings
with respect to full DFT calculations are not compromised as evidenced by
numerical benchmark data.

\vspace*{0.5cm}
\end{abstract}

\section{Introduction}
\label{sec:intro}
Density functional theory (DFT) is currently the de-facto standard in
computational chemistry and condensed matter physics \cite{Burke2012}. Due to
algorithmic improvements and the availability of high
performance computing architectures, simulations with several thousands of atoms are
becoming possible \cite{Bowler2002}. This development paves the way for in-silico combinatorial
chemistry and materials science \cite{Xiang1995,Jain2013}. 

Notwithstanding, more approximate methods like the well established density functional
based tight-binding method (DFTB) still have their merits \cite{Elstner2014}. The search
for global minima on complex potential energy surfaces or the need for accurate phase space sampling in
molecular dynamics simulations are just two examples in this context \cite{Cui2014}. On the methodological level, DFTB has been extended in various ways
during the last years. Such developments include the inclusion of
spin-orbit interactions and non-collinear spin \cite{Kohler2007}, van-der-Waals
corrections \cite{Elstner2001} and hybrid QM/MM schemes \cite{Han2000,Cui2001}, as well as the description of
electronic excited states using the time-dependent formalism TD-DFTB \cite{Niehaus2001a,Niehaus2009,Dominguez2013,Ruger2014}.

Surprisingly, one of the most trivial steps in a DFT calculation, the
change of the employed exchange-correlation (XC) functional, is
comparatively cumbersome in DFTB. This is because DFTB relies on
precomputed matrix elements of the Hamiltonian that are transformed to
the molecular frame by Slater-Koster rules \cite{Slater1954}. In addition, a consistent
set of transferable pair
potentials must be created by combined DFT/DFTB calculations on
reference structures. While these steps are labor-intensive but
straightforward for local and semi-local XC functionals, additional
complications arise for hybrid and long-range corrected (LC) functionals
that involve a fraction of Hartree-Fock exchange. In this case even
the validity of the Slater-Koster rules is far from obvious and the
quality of typical approximations in the DFTB framework like the
neglect of three-center and crystal field terms must be newly
assessed. 

In previous work, the formal foundations of DFTB with LC functionals
were already outlined \cite{Niehaus2012}. In this contribution we
provide details of an efficient implementation of these ideas and
benchmark the accuracy of the new scheme. The interest in LC
functionals arises due to their inherent reduction of the notorious
self-interaction error in DFT. The resulting benefits of these
functionals are well documented
\cite{Gill1996,Savin1996,savin97,Yanai200451,hirao01,Baer2010,Kronik2014}. 
They yield the correct
asymptotic -1/r behavior of the generalized Kohn-Sham potential which
allows for the extraction of accurate ionization potentials and electron
affinities (i.e., quasi-particle energies) from a single N-particle
computation. 
In addition, the localization of the electron density in
extended systems - the typical realm of DFTB applications - is
strongly improved with respect to local functionals. We demonstrate
that these desirable properties are also operational in the DFTB method
which invokes additional approximations beyond the choice of the XC
functional.  

After a specification of the XC functional used in this work
in Section \ref{func}, we briefly summarize the results obtained in
Ref.\ \citenum{Niehaus2012}
before we provide a detailed account of our implementation. Section \ref{results}
is devoted to the results of the new approach and contains a
benchmark of fundamental gaps for a series of organic molecules. In
addition we discuss the molecular response to electric fields
(Section \ref{response}) and
comment shortly on the benefits of the new method for biological
systems in Section \ref{proteins}.

\section{Theory}
\label{sec:theory}
\subsection{Choice of the exchange-correlation functional}
\label{func}
In this work we use a functional that is based on the work of Baer, Neuhauser and Livshits \cite{BaerNeuhauser2005,LivshitsBaer2007}. 
Baer and Neuhauser use the adiabatic connection with a descreened electron-electron interaction $v=[1-\exp{(-\omega r)}]/r$ and derive an approximate 
form of the exchange-correlation functional. 
This form depends only on the
range-separation parameter $\omega$ and provides the correct
long-range behaviour of the potential:  
\begin{equation}
\label{eq:xcfunctotal}
  E_{xc} = E^{\omega,DFT}_{xc} + E^{\omega,HF}_{x}.
\end{equation}
The second term is given by  
\begin{equation}
\label{eq:xcfunchf}
  E^{\omega,HF}_{x} = -\sum^{N/2}_{i,j} \int\int^\prime \psi_i(\mathbf{r}) \psi_j(\mathbf{r}) 
\frac{1 - e^{-\omega |\mathbf{r} - \mathbf{r'}|}}{|\mathbf{r} - \mathbf{r'}|} \psi_i(\mathbf{r'}) \psi_j(\mathbf{r'}) d\mathbf{r'} d\mathbf{r},
\end{equation}
where Kohn-Sham molecular orbitals are denoted by $\psi_i$. The
expression in Eq.\ \ref{eq:xcfunchf} approaches the exact Hartree-Fock (HF) exchange energy for large
inter-electronic distance and is responsible for the reduction of the
self-interaction error.  

The explicit form of the DFT exchange-correlation functional $E^{\omega,DFT}_{xc}$ has still to be approximated. 
Livshits and Baer suggested to use a correlation functional in the
generalized-gradient approximation (GGA) together with short-range
exchange in the local density approximation (LDA) according to   
\begin{equation}
\label{eq:xcfunc2}
  E^{\omega,DFT}_{xc} = E^{GGA}_{c} + \alpha E^{\omega,LDA}_x,
\end{equation}
with some empirically determined parameter $0 \leq \alpha \leq 1$. 
The parameter set  $\{\alpha,\omega\}$ was subsequently optimized by fitting to thermochemical data. 
Similar empirical schemes are widely used to calibrate the whole plethora of different hybrid exchange-correlation functionals\cite{Vydrov2006,Rohrdanz2009,Gill1996,Yanai200451,Akinaga2008348}. 
The aim of this work, however, is a proof of concept and we are
interested in the qualitative improvement over the
traditional DFTB scheme with respect to the failures due to the delocalization problem.
Because of this we do not attempt any fine-tuning of the employed
functional in the present work and simply use the PBE \cite{PBE96} correlation functional and $\alpha=1$ in Eq.\ \ref{eq:xcfunc2}. 

For the LDA short-range DFT functional $E^{\omega,LDA}_x$, the exchange energy of the homogeneous electron gas with Yukawa interaction is used:\cite{Robinson1962,SavinFlad1995,Seth2012}
\begin{align}
\label{eq:xcfuncdft}
  E^{\omega,LDA}_{x}[\rho] &= -\frac{3}{4}\left(\frac{3}{\pi}\right)^{1/3} \int \rho^{4/3}(\mathbf{r}) P(a(\mathbf{r}))\;d\mathbf{r} \\ 
P(a) &=  1 - \frac{8}{3} a\Bigg(\arctan \frac{1}{a} + \frac{1}{4} 
- \frac{a}{4}\left(a^2 + 3\right)\ln\left(1 + \frac{1}{a^2}\right)\Bigg)  \\\label{eq:xcfuncdft_c}
a &= \frac{\omega}{2 k},\;\; k = (3\pi^2\rho)^{1/3}.
\end{align}
Results with this choice of xc functional will be termed LC-DFTB in
the following. 
\subsection{Total energy and Hamiltonian}
In this section we provide a brief overview of the LC-DFTB method,
following  Ref.\ \citenum{Niehaus2012}, together with an
account of our practical implementation.
The method is based on the linear combination of atomic orbitals (LCAO), where
the single particle molecular orbitals $\psi_i$ are expanded into a set of atom-centered basis functions  $\lbrace \phi_\mu \rbrace$: 
\begin{align}
 \psi_i(\mathbf{r}) = \sum_\mu c_{\mu,i} \phi_\mu(\mathbf{r}).
\end{align}
As in the traditional DFTB method, a minimal valence-only basis set is
obtained from pseudo-atomic DFT calculations with the additional
confinement potential $V_{\text{conf}}(r)=(r/r_0)^2$. The pseudo-atoms
are considered to be spherically symmetric by equally distributing
electrons over degenerate valence shells. In this study
the functional is long-range corrected and defined by
Eqs.\ \ref{eq:xcfunctotal} to \ref{eq:xcfuncdft_c}.
The compression radius $r_0$ is usually taken to be proportional to the covalent radius of the respective atomic species. 
In the whole scheme, there are two confinement radii per element:
$r^{\text{basis}}_0$ amounts to the basis set compression and 
$r^{\text{density}}_0$ is used for the compression of the atomic densities in the initial guess.\cite{Elstner2014} 
For LC-DFTB we use the same confinement radii as previously reported
for the mio-1-1 set,\cite{Elstner1998,Niehaus2001} which are summarized in
Table \ref{tab:covradii}. We note that the basis set for sulfur
features additional polarization functions\cite{Niehaus2001}.

\begin{table}[t]
  \centering
  \begin{tabular}{lrrrrrr}
\hline
\hline
  element   & \multicolumn{3}{c}{$r^{\text{basis}}_0$} & \multicolumn{3}{c}{$r^{\text{density}}_0$} \\
\hline
        & s & p & d & s & p & d \\
\hline
    H   & 3.0  &      &       & 2.5  &  & \\
    C   & 2.7  & 2.7  &   & 14.0  & 14.0 & \\
    N   & 2.7  & 2.7  &   & 14.0  & 14.0 & \\
    O   & 2.3  & 2.3  &   &  9.0  &  9.0 & \\
    S   & 3.8  & 3.8  &  4.4  &  9.0  &  9.0 & 9.0 \\
\hline
\hline
  \end{tabular}
  \caption{The compression radii [a$_0$] for the elements used in this work.}
  \label{tab:covradii}
\end{table}

The non-interacting reduced density matrix for the closed-shell case
is given by
\begin{align}
  \rho(\mathbf{r},\mathbf{r'}) = 2 \sum^{N/2}_{i=1} \psi_i(\mathbf{r})\psi_i(\mathbf{r'}) = \sum_{\mu\nu} \underbrace{\left[2\sum^{N/2}_{i=1} n_i c_{\mu,i} c_{\nu,i}\right]}_{P_{\mu\nu}} \phi_\mu(\mathbf{r})\phi_\nu(\mathbf{r'}),
\end{align}
which defines the density matrix $(\mathbf{P})_{\mu\nu} = P_{\mu\nu}$ in
the atomic orbital (AO) representation. 

With these notations and the particular choice of $E^{\omega,\text{DFT}}_{\text{xc}}$ we are ready to write the 
total energy of the closed-shell LC-DFT in the AO basis 
\begin{align}
  E &= \sum_{\mu\nu} P_{\mu\nu} h_{\mu\nu} + \frac{1}{2}\sum_{\mu\nu\alpha\beta} P_{\mu\nu} P_{\alpha\beta} (\mu\nu|\alpha\beta) \\
    &- \frac{1}{4} \sum_{\mu\nu\alpha\beta} P_{\mu\nu} P_{\alpha\beta} (\mu\alpha|\beta\nu)^{\text{lr}} + E^{\omega,\text{DFT}}_{\text{xc}}[\rho],
\end{align}
where single particle Hamiltonian $h_{\mu\nu}$ and two-electron
repulsion integrals are given by 
\begin{align}
  h_{\mu\nu} &= \int \phi_\mu(\mathbf{r})\left(-\frac{1}{2}\nabla^2 + v^{\text{ext}}(\mathbf{r})\right)\phi_\nu(\mathbf{r}) d\mathbf{r}  \\
(\mu\nu|\alpha\beta) &= \int\int \phi_\mu(\mathbf{r})\phi_\nu(\mathbf{r})\frac{1}{|\mathbf{r}-\mathbf{r'}|} \phi_\alpha(\mathbf{r'})\phi_\beta(\mathbf{r'}) d\mathbf{r'} d\mathbf{r} \\
(\mu\nu|\alpha\beta)^{\text{lr}} &= \int\int \phi_\mu(\mathbf{r})\phi_\nu(\mathbf{r})\frac{1-e^{-\omega|\mathbf{r}-\mathbf{r'}|}}{|\mathbf{r}-\mathbf{r'}|} \phi_\alpha(\mathbf{r'})\phi_\beta(\mathbf{r'}) d\mathbf{r'} d\mathbf{r};
\end{align}
here $v^{\text{ext}}$ denotes the external potential. 
Next, the density matrix is decomposed as $\mathbf{P} = \mathbf{P}_0 +
\Delta \mathbf{P}$, with some reference density matrix or initial guess $\mathbf{P}_0$. 
Such a decomposition is common for direct SCF approaches where the Hamiltonian is constructed iteratively \cite{Haeser1989}. 
We choose the reference density matrix to be the superposition of
atomic density matrices $\mathbf{P}_0 = \sum_A \mathbf{P}_A$, which
are obtained from the previous atomic LC-DFT calculations for each
element. This choice parallels on the one hand the protocol in
standard DFTB, on the other hand it is widely used as the initial guess in quantum chemistry codes.

We now expand the local part of the exchange-correlation functional around the reference density matrix up to second order in $\Delta \mathbf{P}$:
\begin{align}
\label{expand}
  E^{\omega,\text{DFT}}_{\text{xc}}[\rho] &= E^{\omega,\text{DFT}}_{\text{xc}}[\rho_0] + \int \frac{\delta E^{\omega,\text{DFT}}_{\text{xc}}}{\delta \rho(\mathbf{r})}\Big|_{\rho_0} \delta \rho(\mathbf{r}) d\mathbf{r} \nonumber  \\
&+ \frac{1}{2} \int \frac{\delta^2 E^{\omega,\text{DFT}}_{\text{xc}}}{\delta \rho(\mathbf{r})
  \delta \rho(\mathbf{r'}) }\Big|_{\rho_0} \delta \rho(\mathbf{r}) \delta \rho(\mathbf{r'}) d\mathbf{r}d\mathbf{r'} + \mathcal{O}(\delta \rho^3) \nonumber \\
&= E^{\omega,\text{DFT}}_{\text{xc}}[\rho_0] + \sum_{\mu\nu} \Delta P_{\mu\nu} v^{\omega,\text{xc}}_{\mu\nu}[\rho_0] \nonumber \\ 
&+ \frac{1}{2}\sum_{\mu\nu\alpha\beta} \Delta P_{\mu\nu} \Delta P_{\alpha\beta} f^{\omega,\text{xc}}_{\mu\nu\alpha\beta}[\rho_0] + \mathcal{O}(\delta \rho^3).
\end{align}
First and second derivatives of the functional in the AO basis are
denoted here as $v^{\omega,\text{xc}}_{\mu\nu}$ and $f^{\omega,\text{xc}}_{\mu\nu}$, respectively.
Eq.\ \ref{expand} represents the first approximation in LC-DFTB, which amounts to linearization of the local exchange-correlation potential. 
Inserting this into the total energy expression and following the
procedure in Ref.\ \citenum{Elstner1998}, we rearrange the terms such that 
\begin{align}
\label{etot}
  E &= \sum_{\mu\nu} P_{\mu\nu} H^0_{\mu\nu} + \frac{1}{2}\sum_{\mu\nu\alpha\beta} \Delta P_{\mu\nu} \Delta P_{\alpha\beta} \left[ (\mu\nu|\alpha\beta) + f^{\omega,\text{xc}}_{\mu\nu\alpha\beta}[\rho_0] \right] \nonumber \\
    &- \frac{1}{4} \sum_{\mu\nu\alpha\beta} \Delta P_{\mu\nu} \Delta P_{\alpha\beta} (\mu\alpha|\beta\nu)^{\text{lr}} + E_{\text{rep}}, 
\end{align}
where the last term, the repulsive energy $E_{\text{rep}}$ \cite{Elstner2014},  depends only on the reference density. 
The second approximation is to replace the repulsive energy by a sum of fast decaying pair potentials as it is the case in standard DFTB \cite{Elstner1998}.
This is a reasonable approximation also in the present range-separated
formalism as shown in Ref.\ \citenum{Niehaus2012}. 

The Hamiltonian $H^0_{\mu\nu}$, evaluated at the reference density
\begin{align}
  H^0_{\mu\nu} &= h_{\mu\nu} + \sum_{\alpha\beta} P^0_{\alpha\beta} (\mu\nu|\alpha\beta) \nonumber \\
 &- \frac{1}{2}\sum_{\alpha\beta} P^0_{\alpha\beta} (\mu\alpha|\beta\nu)^{\text{lr}} + v^{\omega,\text{xc}}_{\mu\nu}[\rho_0]
\end{align}
is treated in the two-center approximation
\begin{align}
\label{h0}
  H^0_{\mu\nu} =
  \begin{cases}
    \epsilon^{\text{free}}_{\mu} & \mu = \nu \\
    H^0_{\mu\nu}[\mathbf{P}_A + \mathbf{P}_B] & \mu\in A,\;\; \nu\in B \\
    0 & \text{else},
  \end{cases}
\end{align}
where the density matrix of the dimer $\mathbf{P}_A + \mathbf{P}_B$ is
constructed with the density compression radius
$r^{\text{density}}_0$. As already mentioned, the basis functions
$\phi_\mu$ stem from a LC-DFT calculation with the compression radius
$r^{\text{basis}}_0$ and are used to construct both the zeroth-order
Hamiltonian matrix elements $H^0_{\mu\nu}$ and the overlap matrix
$S_{\mu\nu} = \int
\phi_\mu(\mathbf{r})\phi_\nu(\mathbf{r})\;d{\mathbf{r}}$. Thanks to
the two-center approximation, matrix elements for all possible
geometries can be constructed from 
a small set of high-symmetry integrals according to Slater-Koster rules \cite{Slater1954}. 
These parameters are thus tabulated as a function of inter-atomic
distance $R_{AB}=
|\mathbf{R}_{A}-\mathbf{R}_B|$. 

Next we approximate the two-electron integrals using the Mulliken approximation 
\begin{align}
\label{mull}
  \phi_\mu(\mathbf{r})\phi_\nu(\mathbf{r}) \approx \frac{1}{2} S_{\mu\nu} \left( |\phi_\mu(\mathbf{r})|^2 + |\phi_\nu(\mathbf{r})|^2\right).
\end{align}
This allows for reduction of four-center integrals by a sum of two-center integrals $\gamma_{\mu\nu}$ 
\begin{align}
  (\mu\nu|\alpha\beta)^{\text{lr}} \approx& \frac{1}{4}S_{\mu\nu} S_{\alpha\beta} \left[ (\mu\mu|\alpha\alpha)^{\text{lr}}  + (\mu\mu|\beta\beta)^{\text{lr}} + (\nu\nu|\alpha\alpha)^{\text{lr}} + (\nu\nu|\beta\beta)^{\text{lr}} \right] \\
  =& \frac{1}{4}S_{\mu\nu} S_{\alpha\beta} \left[ \gamma^{\text{lr}}_{\mu\alpha} + \gamma^{\text{lr}}_{\mu\beta} + \gamma^{\text{lr}}_{\nu\alpha} + \gamma^{\text{lr}}_{\nu\beta} \right].
\end{align}
In the spirit of the original DFTB method,\cite{Elstner1998} these integrals are then parametrized as 
 \begin{align}
\label{eq:gammalr}
 \gamma^{\text{lr}}_{\alpha\beta} &= \gamma^{\text{lr}}_{AB}(R_{AB}) = \frac{\tau^3_A \tau^3_B}{(8\pi)^2}
\int\int e^{-\tau_A |\mathbf{r}_{1}-\mathbf{R}_A|}\frac{1-e^{-\omega |\mathbf{r}_{1} - \mathbf{r}_{2}|}}{|\mathbf{r}_{1} - \mathbf{r}_{2}|} e^{-\tau_B |\mathbf{r}_{2}-\mathbf{R}_B|} \;d\mathbf{r}_1 d\mathbf{r}_2,
 \end{align}
where $\alpha\in A,\; \beta\in B$, and $\tau_A$ is some atom specific
decay constant, which has to be defined. In a similar fashion the full
range two-electron integrals are simplified to
 \begin{align}
\label{eq:gammafr}
 \gamma^{\text{fr}}_{\alpha\beta} &= \gamma^{\text{fr}}_{AB}(R_{AB}) = \frac{\tau^3_A \tau^3_B}{(8\pi)^2}\
\int\int e^{-\tau_A |\mathbf{r}_{1}-\mathbf{R}_A|}\left[\frac{1}{|\mathbf{r}_{1} - \mathbf{r}_{2}|} + f^{\omega,\text{xc}}[\rho_0](\mathbf{r}_{1},\mathbf{r}_{2})\right] e^{-\tau_B |\mathbf{r}_{2}-\mathbf{R}_B|} \;d\mathbf{r}_1 d\mathbf{r}_2. 
 \end{align}  

\begin{figure}[t]
  \centering
  \includegraphics[width=0.8\textwidth]{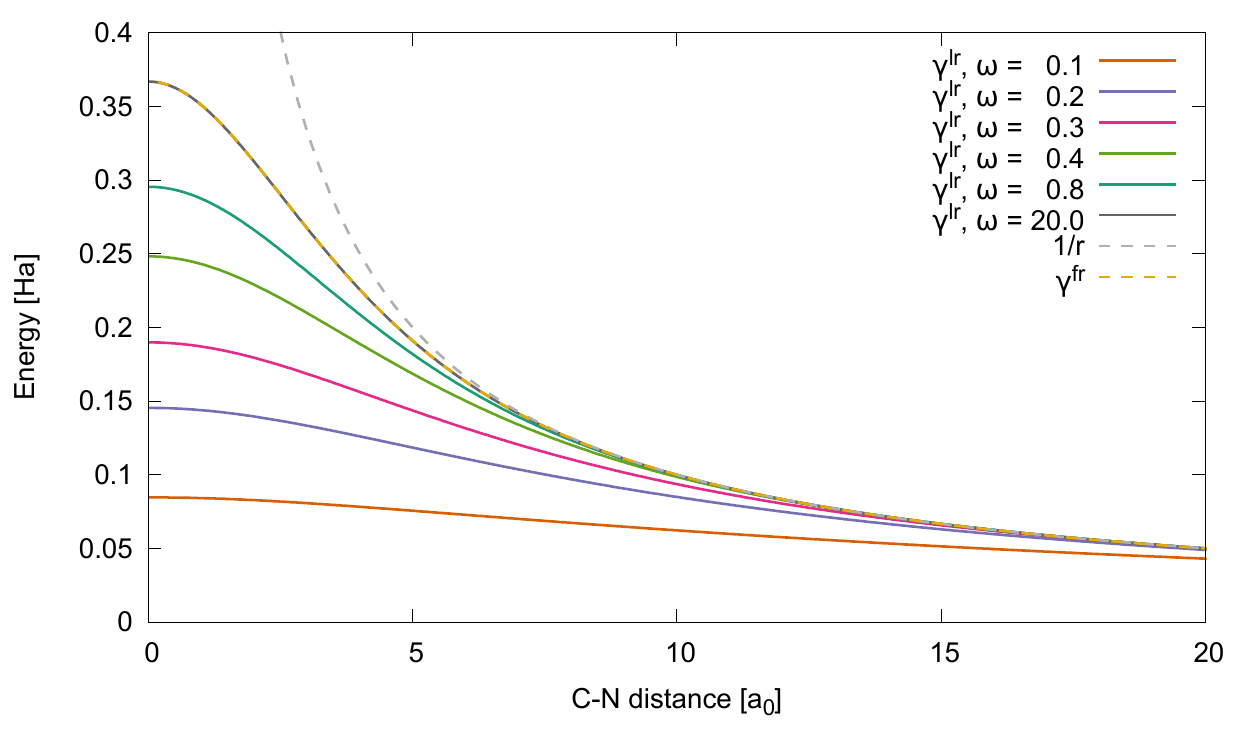}
  \caption{The long-range integral $\gamma^\text{lr}$  for the carbon-nitrogen
    interaction as a function of inter-atomic distance for different
    values of the range-separation parameter $\omega$.
The gray dashed line denotes the $1/r$ limit. 
The full-range $\gamma^{\text{fr}}$ as defined in appendix of Ref.\
\citenum{Elstner1998} is depicted with an orange dashed
line. 
}
  \label{fig:lrgamma}
\end{figure}

To obtain the final prescription for the evaluation of these integrals we follow the reasoning of the original DFTB method.
For the off-site elements $A\neq B$, the contribution due to the xc-kernel $f^{\omega,\text{\text{xc}}}$ is assumed to vanish. 
Thus the full-range $\gamma$-integral reduces to a two-center Coulomb integral over simple spherically symmetric Slater type functions. 
For this case an analytical result is available\cite{Elstner1998},
which is also used in the present scheme. Exchange-correlation contributions are fully taken into account for
the on-site case $A = B$, as shown below. To evaluate the long-range $\gamma$-integrals, 
we extend the analytical formula of Elstner et al. \cite{Elstner1998} to the case of Yukawa interaction
(see appendix \ref{sec:apphubbard} for details). As example, we plot $\gamma^{\text{lr}}_{\text{C-N}}$ for the
carbon-nitrogen interaction as a function of inter-atomic distance for
some values of the range-separation parameter $\omega$
in Figure \ref{fig:lrgamma}.
Additionally, the full-range $\gamma^{\text{fr}}$ evaluated with the
analytical formula from Ref.\ \citenum{Elstner1998} is also depicted (orange dashed line). 

For the on-site elements $A=B$, the integrals have the form (details
are given in appendix \ref{sec:apphubbard})
\begin{align}
 \lim_{R_{AB}\rightarrow 0} \gamma^{\text{lr}}_{AB}(R_{AB}) &= \frac{5}{16}\tau_A - 
\frac{\tau^8_A}{(\tau^2_A - \omega^2)^4} \nonumber \\ &\times \left[\frac{5\tau^6_A + 15\tau^4_A\omega^2 - 5\tau^2_A\omega^4 + \omega^6}{16\tau^5_A} - \omega\right] \\
\lim_{R_{AB}\rightarrow 0} \gamma^{\text{fr}}_{AB}(R_{AB}) &= \frac{5}{16}\tau_A.
\end{align}
For each element, the Hubbard parameter ($U = \partial^2 E / \partial
  n^2$) is defined as the second derivative of the energy with respect to
the occupation of the highest occupied atomic orbital. The decay constants $\tau_A$ may therefore be determined by the requirement
that LC-DFT and LC-DFTB yield the same Hubbard value for each species.    
In the presence of a Fock term this relation is different from that in
traditional DFTB ($U_A = 5/16 \tau_A$). 
The Hubbard parameter of an atom in LC-DFTB is instead given by 

\begin{align}
\label{hub}
U^{\text{LC-DFTB}}_A &= \gamma^{\text{fr}}_{AA} - \frac{1}{2(2l+1)}\gamma^{\text{lr}}_{AA}  
  = \frac{5}{16}\tau_A\Bigg[1-\frac{1}{2(2l+1)} \nonumber \\
&\times \Big(1-\frac{\tau_A^8+3\tau_A^6\omega^2 -\tau_A^4\omega^4 + 0.2\omega^6\tau_A^2 - 3.2\tau_A^7\omega}{(\tau_A^2-\omega^2)^4}\Big)\Bigg],
\end{align}
where $l$ is the angular momentum of the highest atomic
orbital (see appendix \ref{sec:hubpar} for details). 
Evaluating $U_A$ using first principles LC-DFT for a given
range-separation parameter $\omega$, Eq.\
\ref{hub} is solved numerically for $\tau_A$.

Having defined all elements of the method we proceed by applying the
variational principle to the total energy in Eq.\ \ref{etot}. This
yields the Hamiltonian 
\begin{align}
\label{eq:lcdftbhamiltonian}
  H_{\mu\nu} &= H^0_{\mu\nu} + \frac{1}{4}\sum_{\alpha\beta} \Delta P_{\alpha\beta} S_{\mu\nu} S_{\alpha\beta} \left(\gamma^{\text{fr}}_{\mu\alpha}
 + \gamma^{\text{fr}}_{\mu\beta} + \gamma^{\text{fr}}_{\nu\alpha} + \gamma^{\text{fr}}_{\nu\beta}\right) \nonumber \\
&-\frac{1}{8}\sum_{\alpha\beta} \Delta P_{\alpha\beta} S_{\mu\alpha} S_{\beta\nu} 
\left(\gamma^{\text{lr}}_{\mu\beta} + \gamma^{\text{lr}}_{\mu\nu} + \gamma^{\text{lr}}_{\alpha\beta} + \gamma^{\text{lr}}_{\alpha\nu} \right),
\end{align}
and the generalized Kohn-Sham equations 
\begin{align}
 \sum_{\nu} H_{\mu\nu} c_{\nu,i} = \epsilon_i \sum_{\nu} S_{\mu\nu} c_{\nu,i}.
\end{align}
In contrast to the original
DFTB scheme, the full density matrix needs to be evaluated self-consistently
rather than just the Mulliken charges.

\subsection{Implementation notes}
\label{impl}
The scheme has been implemented in the development version of the DFTB+
code.\cite{Aradi2007a} With respect to the original DFTB method, two
significant changes are necessary. First, the zeroth-order Hamiltonian
(Eq.\ \ref{h0})
now also includes a contribution due to screened Hartree-Fock
exchange. The necessary adaption of the numerical integration routines
is outlined in appendix \ref{app:integrals}. Since the $H^0$ and $S$
matrix elements are precomputed there is no overhead involved in actual calculations.   

The second change is related to the presence of an exchange term in
the Fock matrix, given by the second line in
Eq. \ref{eq:lcdftbhamiltonian}. This part is performance-critical,
since a naive implementation would lead to $N^4$ scaling,
despite of the applied Mulliken approximation (Eq.\ \ref{mull}).         

We therefore resort to techniques which are widely used in direct SCF
calculations.\cite{Haeser1989} The Hamiltonian is constructed iteratively 
\begin{align}
\label{iter}
  H(\mathbf{P}^n) = H(\mathbf{P}^{n-1} + \Delta \mathbf{P}) = H(\mathbf{P}^{n-1}) + H(\Delta \mathbf{P}),
\end{align}
such that the Hamiltonian at the $n-$th iteration is the sum of the
Hamiltonian at the previous iteration plus a
correction. Here the density matrix at $n-$th iteration is labeled $\mathbf{P}^n$
and $\Delta \mathbf{P}$ denotes the difference
  between density matrices at different iterations of the
  self-consistent cycle, not the difference between ${\bf P}$ and ${\bf
    P_0}$ as in Eq.\  \ref{expand}.

We then exploit the fact that the matrix elements
$H(\Delta\mathbf{P})$ get smaller upon approaching convergence. In
order to make efficient use of integral prescreening techniques, the
exchange matrix in Eq. \ref{eq:lcdftbhamiltonian} is rewritten in the following form
\begin{align}
\label{kexch}
  K_{\mu\nu} &= -\frac{1}{8}\sum_{\alpha\beta} \Delta P^n_{\alpha\beta} S_{\mu\alpha} S_{\beta\nu} 
\left(\gamma^{\text{lr}}_{\mu\beta} + \gamma^{\text{lr}}_{\mu\nu} + \gamma^{\text{lr}}_{\alpha\beta} + \gamma^{\text{lr}}_{\alpha\nu} \right) \\
&= -\frac{1}{8}\sum_{AB}\underbrace{\left[\gamma^{\text{lr}}_{CB} + \gamma^{\text{lr}}_{AB} + \gamma^{\text{lr}}_{CD} + \gamma^{\text{lr}}_{AD}\right]}_{\Gamma_{ABCD}} 
\sum_{\alpha\in A}\sum_{\beta\in B} S_{\mu\alpha} \Delta P^n_{\alpha\beta} S_{\beta\nu} \\ 
&= -\frac{1}{8}\sum_{AB}\Gamma_{ABCD} Q^{\mu\nu}_{AB}.
\end{align}
For each Hamiltonian sub-block (C,D) and atom pair (A,B), an estimate
for the quantities $Q^{\mu\nu}_{AB},\mu\in C,\nu\in D$ is given by 
 \begin{align}
\label{eq:estimate}
   Q^{\mu\nu}_{AB} \leq \sum_{\alpha\in A} \sum_{\beta\in B}
   |S_{\mu\alpha}||\Delta P^n_{\alpha\beta}||S_{\beta\nu}|  \leq
   \overline{S}_{BD} \overline{S}_{AC}  \overline{\Delta P^n}
   \left(\sum_{\alpha\in A} \sum_{\beta\in B} 1\right),
 \end{align}
where $\overline{S}_{AB} = \max\limits_{\alpha\in A,\beta\in
  B}(|S_{\alpha\beta}|)$ and $\overline{\Delta P^n}=\max(|\Delta
P^n_{\alpha\beta}|)$. 
Imposing the cutoff criterion 
\begin{align}
  \overline{S}_{BD} \overline{S}_{AC} \overline{\Delta P^n} \leq \epsilon_{\text{threshold}},
\end{align}
we decide whether the contribution of the Hamiltonian matrix element is significant and avoid the evaluation
if possible. The double sum on the right side of Eq.\ \ref{eq:estimate} is absorbed in $\epsilon_{\text{threshold}}$.

\begin{figure}[t]
  \centering
  \includegraphics[width=0.8\textwidth]{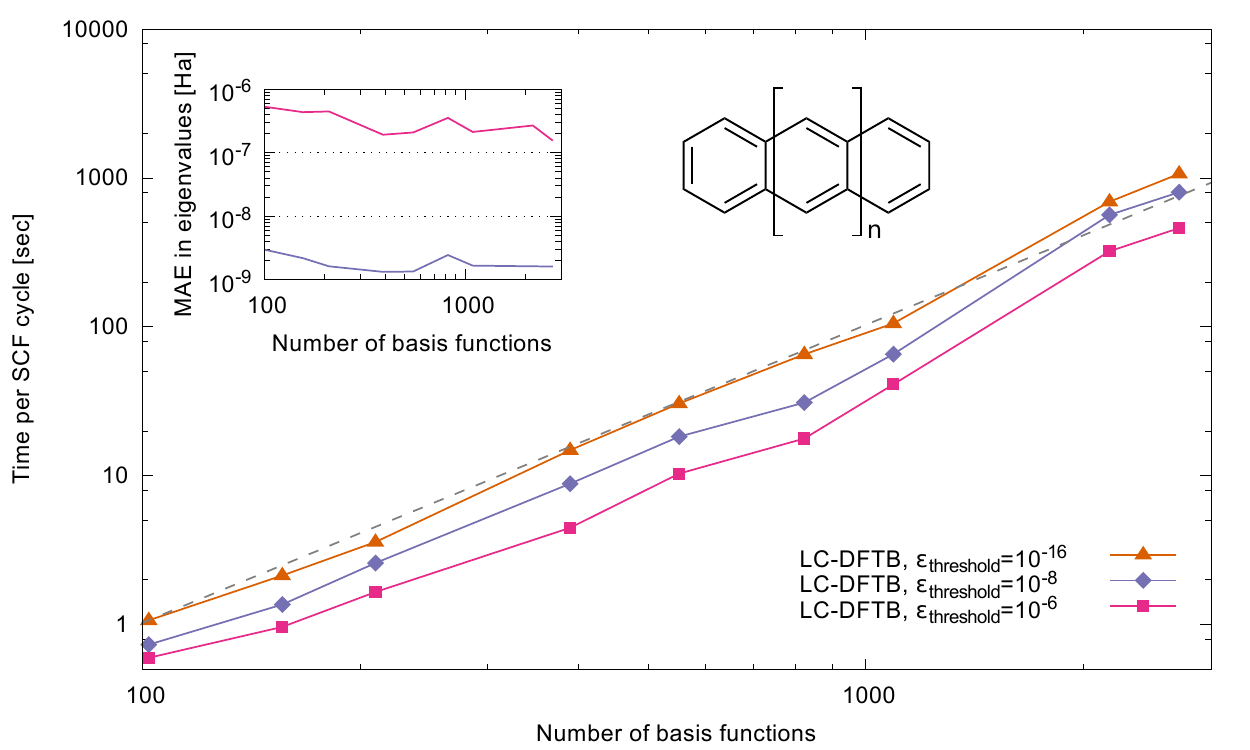}
  \caption{The average time per SCF cycle for the polyacene oligomers
    versus the number of basis functions for the threshold parameters $\epsilon_{\text{threshold}} = 10^{-16}, 10^{-8}, 10^{-6}$. 
The gray dashed line gives an extrapolation for ideal quadratic
scaling. The inset shows the mean absolute error in the eigenvalues in
Hartree for the cases of $\epsilon_{\text{threshold}} = 10^{-8}$ and
$\epsilon_{\text{threshold}} = 10^{-6}$. 
Calculations have been performed on a single core of an Intel Core-i7 CPU. 
The execution time was measured by the Linux {\em time}
utility. 
}
  \label{fig:scaling}
\end{figure}

We analyze the efficiency and accuracy of this prescreening approach
by performing benchmark calculations for the polyacene series
($C_{4n+2}H_{2n+4}$) up to $n=150$. Figure \ref{fig:scaling} depicts the scaling of the method with respect to the basis size. 
The gray dashed line shows the extrapolated quadratic function $t(n) =
c n^2$, with $t(n_0)$ equal to the CPU time at the smallest oligomer
size $n_0=5$ 
and 
$\epsilon_{\text{threshold}} = 10^{-16}$. 

For $\epsilon_{\text{threshold}} = 10^{-16}$, prescreening
and exact evaluation according to Eq.\ \ref{kexch} lead to identical
total energies within machine precision. 
Even with this tight threshold
criterion, the expected quadratic scaling is achieved. 
For the finite threshold values of $\epsilon_{\text{threshold}} =
10^{-8}$ and $\epsilon_{\text{threshold}} = 10^{-6}$ the scaling remains quadratic, however the 
prefactor is reduced by a factor of 2-3. In the inset the mean
absolute error (MAE) of the eigenvalues with respect to the exact evaluation is shown. 
For $\epsilon_{\text{threshold}}=10^{-6}$ this error does not exceed $10^{-6}$ Ha. 

The overall scaling of the method is determined by the calculation
step with steepest scaling. 
Since the Hamiltonian matrix construction scales quadratically in our scheme, the scaling for very large systems should be determined by the 
$\mathcal{O}(N^3)$ behavior of the diagonalizer. 
However, for the tested systems the time spent for diagonalization is negligible compared to the Hamiltonian construction. 
Thus we expect cubic scaling only for much larger system sizes,
especially if effective Divide \& Conquer diagonalization routines are
available like in the  current DFTB+ version.

\section{Results}
\label{results}

After the discussion of the main approximations and computational
efficiency of the LC-DFTB method, we benchmark its predictive
power. At this point we focus on the electronic structure at fixed
geometry to highlight the advantages with respect to the original DFTB
method in the computation of electron removal and addition
energies. This also represents a necessary first step in developing a
time-dependent formalism for long-range corrected functionals in the spirit
of the TD-DFTB scheme.

\subsection{Quasi-particle energies}
\label{sec:quasipart}

 It is well known that KS eigenvalues from local DFT are poor estimates
for quasi-particle energies \cite{Refaely2012,Cohen2012,Baer2010,Cohen2008,Seidl1996,Perdew1985}. 
Especially the eigenvalue of the highest occupied molecular
orbital (HOMO), if interpreted as electron removal energy (ionization potential) underestimates the 
experimental ionization potential (IP) by several eV. Electron affinities estimated from
energy of the lowest unoccupied molecular orbital (LUMO) are likewise
prone to large errors. 
As a consequence the HOMO-LUMO gap is much smaller than the
experimental fundamental gap or the one obtained from 
the GW approximation\cite{Blase2011,Cohen2012,Kronik2012,Refaely2011}.
This challenges the ability of the local DFT to provide useful single-particle picture of the physical systems. 

Two major problems of local DFT have been identified. 
On the one hand, the exponential asymptotic decay of the KS-potential (instead of the correct $-1/r$ behavior) leads to  
underbound electrons and a wrong description of the long-range interaction. 
On the other hand, the correct asymptotics alone is usually not sufficient for the correct prediction of quasi-particle energies\cite{Cohen2012}.
The correct total energy of a system with fractional occupation $N+\delta$, where $N$ is an integer and $0\le \delta \le 1$ is real,
has been shown to exhibit a linear dependence on particle number between $N$ and $N+1$. 
This is the necessary and sufficient condition to obtain the correct IP and fundamental gap. 
The linearity condition has been directly connected to the many-electron self-interaction error (MSIE) exhibited by 
local exchange-correlation functionals\cite{MoriSanchez2006}, which in a real
space picture manifests as a delocalization problem\cite{Cohen2012}.

The LC-DFT restores the correct  asymptotics of the potential and
shows remarkably close agreement with the linearity condition\cite{Baer2010,Cohen2008,Cohen2012,Koerzdoerfer2012}.
As a consequence, at least frontier orbitals in LC-DFT can be interpreted as electron removal energies.
We expect the LC-DFTB method to show similar qualitative improvement compared to DFTB. 

To investigate this point, we choose a set of organic molecules for which experimental ion energetics data is available. 
Among others this set includes a selection of compounds that are
relevant for photovoltaic applications and have been studied in a similar context with LC-DFT\cite{Refaely2011} and GW\cite{Blase2011} methods.
The structural formulas of these molecules can be found in the supporting material.

All geometries have been optimized at the traditional DFTB level with
the mio-1-1 set\cite{Elstner1998,Niehaus2001} and are used by default for both first principles and DFTB/LC-DFTB calculations. 
For the calculations in this work, we choose a value of  $\omega=0.3$ a$_0^{-1}$
for the range-separation parameter unless stated otherwise. 
We found this value to give reasonable results for the prediction of ionization potentials. 
Akinaga and Ten-no showed, that for a range-separation of Yukawa type the optimal value of the range-separation parameter is
usually higher than that for error function based functionals.\cite{Akinaga2008348} 
In their work the optimization of the parameter was carried out on atomization energies of the G2-1 set with the cc-pVTZ basis. 
However, it is so far not clear whether these parameters can be directly transferred to LC-DFTB. 
From our calculations, we found values ranging from $\omega=0.5$ a$^{-1}_0$ to $\omega=0.75$ a$^{-1}_0$, 
as suggested in Ref.\ \citenum{Akinaga2008348}, to be too large for the prediction of accurate ionization potentials. 
In this case LC-DFTB tends to systematically overestimate IPs.
 
We include the
gradient-corrected PBE\cite{PBE96} functional, the hybrid functional B3LYP\cite{Becke1993,Lee1988,VWN1980,Stephens1994} and the
long-range corrected BNL\cite{LivshitsBaer2007} functional in the comparison. In general a
basis set of triple-zeta quality with polarization functions is
employed, while we also use smaller sets for the BNL functional
to estimate the basis set dependence of the results. All first
principles DFT calculations have been performed with the
NWCHEM-6.3 package.\cite{Valiev2010} 

\begin{figure}
  \centering
  \includegraphics[width=0.8\textwidth]{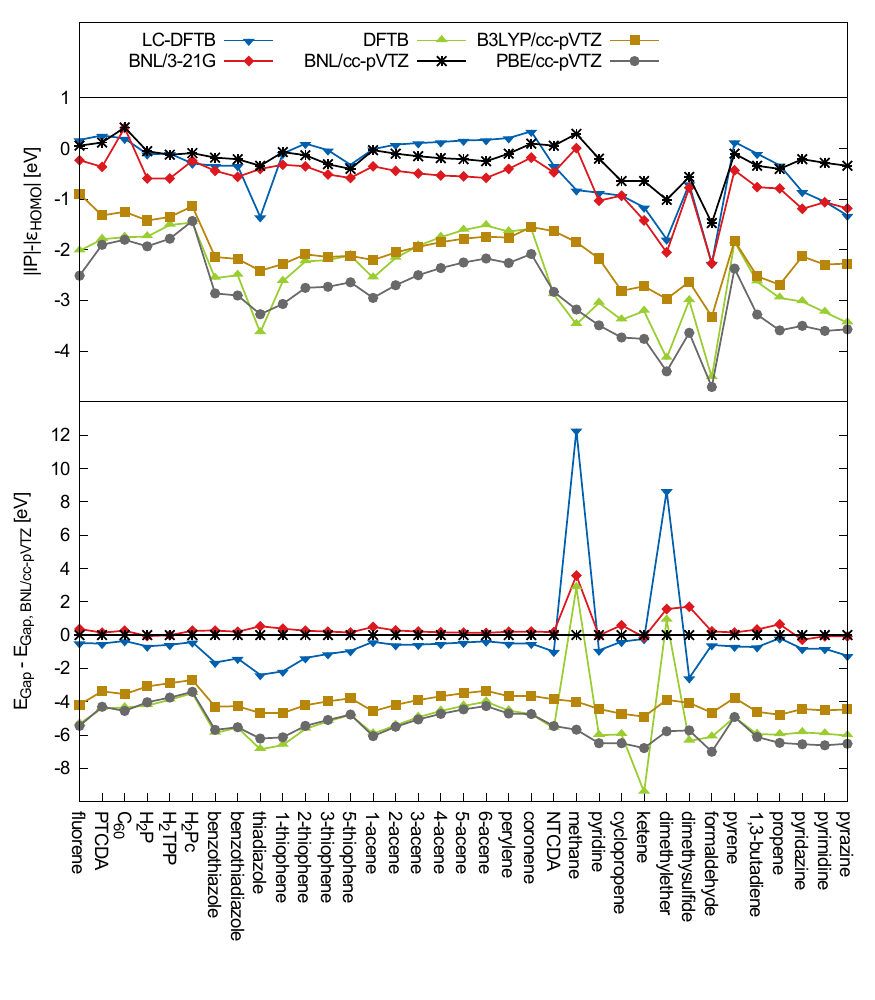}
  \caption{(Top) Deviation of the negative HOMO eigenvalue ($-\epsilon_{\text{HOMO}}$) from the experimental ionization potential for different levels of theory. 
(Bottom) Deviation of the HOMO-LUMO gap from the reference
(BNL/cc-pVTZ). 
}
  \label{fig:combined}
\end{figure}
First we compare the negative of the HOMO eigenvalue to the
experimental ionization potential\cite{nist69} for the mentioned set of molecules. 
The deviation $\Delta=|\text{IP}_\text{exp}| -
|\epsilon_{\text{HOMO}}|$ for LC-DFTB, DFTB, PBE, B3LYP and BNL is shown in
the top part of Figure \ref{fig:combined}. It is found that LC-DFTB is
quantitatively in better agreement with experiment than the standard DFTB and first
principle approaches with the PBE and B3LYP functionals. 
The MAE for LC-DFTB is 0.50 eV compared to BNL/3-21G (0.67 eV),
BNL/cc-pVDZ (0.47 eV), and BNL/cc-pVTZ (0.29 eV). Since the deviations for
the local and hybrid functionals are much larger [B3LYP/cc-pVTZ 
(2.04 eV), PBE/cc-pVTZ (2.87 eV), DFTB (2.50 eV)], we conclude that the assets of
long-range corrected functionals are still visible at the approximated
LC-DFTB level. 

For some molecules, most notably thiadiazole and
methane, the
deviation from both experiment and BNL can however be quite large. This seems
to be an effect of the basis set employed in the LC-DFTB method. 
As already mentioned in Section \ref{sec:theory}, we did not optimize
the compression radii for the new functional and use the parameters of
the mio-1-1 basis set without changes. Furthermore the minimal basis of the DFTB/LC-DFTB method might not provide enough variational flexibility.
This can be deduced from the results for the BNL functional at the 3-21G, cc-pVDZ and cc-pVTZ level. 
Thiadiazole is an exception, since for this compound basis set effects in the first principles
calculations (BNL/3-21G compared to BNL/cc-pVTZ) are small. 
Inspection of Figure \ref{fig:combined} reveals that the results of the
LC-DFTB method
are in general comparable to LC-DFT results with small double-zeta basis
(3-21G). The new scheme clearly outperforms first principles DFT calculations based on
the PBE and B3LYP functionals for the description of ionization potentials.

Next, we investigate fundamental band gaps. 
Since the experimental data for electron affinities is in general not
available, we use BNL/cc-pVTZ results as reference. Deviations of the
fundamental gaps from this reference are depicted in the bottom part of Figure \ref{fig:combined}. 
The MAE deviation for LC-DFTB is 1.36 eV, compared to DFTB (5.06 eV),  BNL/3-21G (0.41 eV), BNL/cc-pVDZ (0.15 eV), PBE/cc-pVTZ (5.29 eV), and B3LYP/cc-pVTZ (3.92 eV). 
For the case of methane and dimethylether rather large deviations of 12.30 eV 
and 8.66 eV, respectively, are found for LC-DFTB. Again we assign
these failures to the minimal basis set, since   
BNL/3-21G shows qualitatively similar, although much smaller,
deviations of 3.58 eV and 1.57 eV for these molecules.
In line with this we note that dimethylsulfide, which is essentially dimethylether with oxygen being replaced by sulfur, 
shows a much smaller error. This can be attributed to the fact that the sulfur in present parametrization contains additional polarization functions (d-orbitals). 
Inclusion of polarization functions for oxygen and nitrogen might
thus reduce the mentioned problem with a moderate  loss of computational efficiency. 

\begin{figure}
  \centering
  \includegraphics[width=0.8\textwidth]{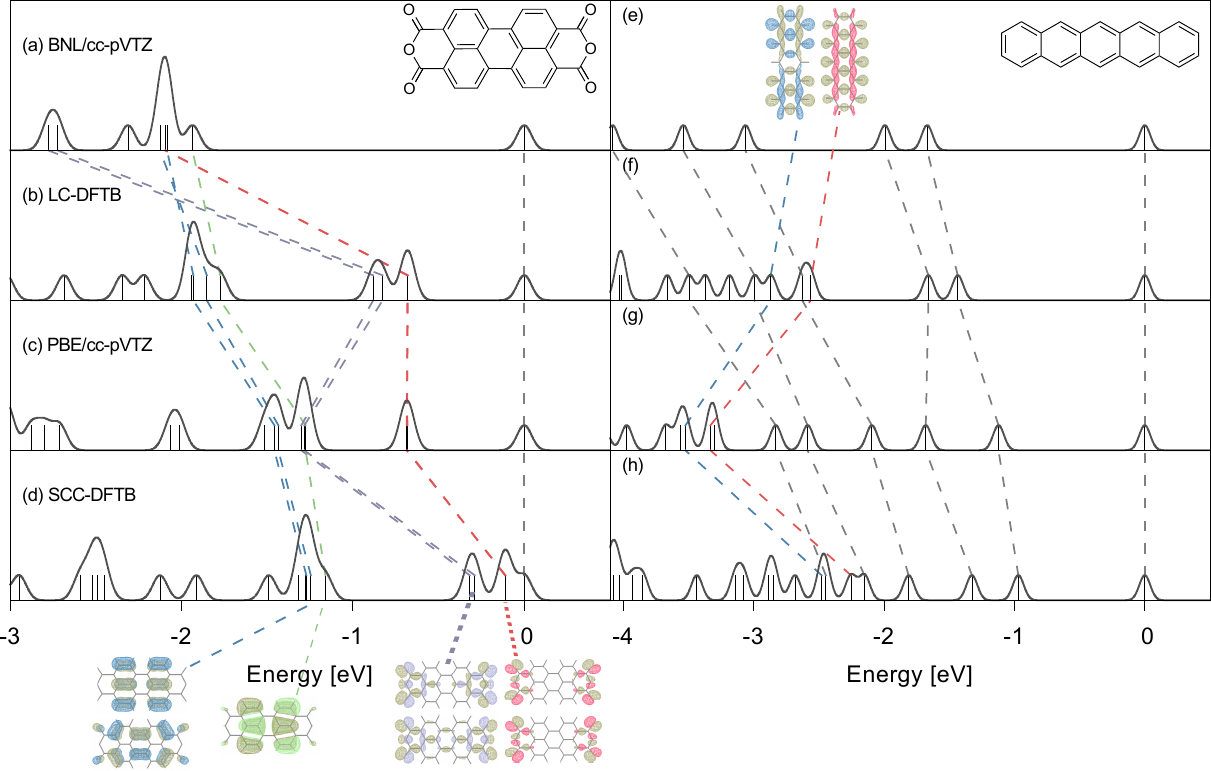}
  \caption{Broadened density of states for PTCDA and pentacene
    molecules from LC-DFTB and DFTB compared to BNL/cc-pVTZ and PBE/cc-pVTZ
    results. The HOMO level has been shifted to the zero of energy for
    all methods.}
  \label{fig:levels}
\end{figure}

We now proceed by investigating the electronic structure beyond the frontier molecular orbitals 
which is experimentally accessible by photoemission spectroscopy.  
Here successful applications of theoretical methods such as GW or
hybrid DFT have been reported, whereas local DFT exhibited serious flaws\cite{Dori2006,Marom2012,Sharifzadeh2012,Kronik2014}. 
The deficiencies of local DFT, which could be attributed to the self-interaction error, seem to be partially cured by  
long-range corrected
functionals\cite{Refaely2012,Koerzdoerfer2012,Dori2006}. The eigenvalue spectrum from standard DFTB, LC-DFTB, PBE/cc-pVTZ and BNL/cc-pVTZ theories for 
3,4,9,10-perylene-tetracarboxylic-dianhydride (PTCDA) and pentacene (5-acene) molecules is presented in Figure \ref{fig:levels}. 
All spectra are rigidly shifted such that HOMO position is at 0 eV. 
We use a simple gaussian broadening profile with the full width at half minimum of 0.1 eV to mimic the experimental resolution and broadening. 
For this study geometries optimized at the B3LYP/cc-pVTZ level have been used.

The experimental photoemission spectrum of the PTCDA 
molecule is characterized by the large gap of 1.5 eV between the first and second peaks, 
where the second peak appears at energies relative to the HOMO between -1.5 eV and -2.1 eV\cite{Sauther2009}. 
Long-range corrected functionals and GW are usually capable to reproduce these features qualitatively.\cite{Kronik2014} 
We confirm this again with our BNL/cc-pVTZ calculation, see part a) of Figure \ref{fig:levels}. 
The spectrum is comparable to the standard LC-$\omega$PBE result of Ref.\ \citenum{Koerzdoerfer2012}.
Since PBE provides poor results as has been already discussed in Refs.\ \citenum{Koerzdoerfer2012,Refaely2012}, 
we do not expect a better performance for DFTB, which is essentially
an approximate DFT with the PBE functional. As can be clearly seen
from part c) and d) of Figure \ref{fig:levels}, the DFTB spectrum
differs even more strongly from BNL/cc-pVTZ than PBE.

We keep this fact in mind and proceed to the LC-DFTB results.  
We observe again a major difference to the BNL/cc-pVTZ spectrum.
The LC-DFTB shows four $\sigma$-orbitals, mostly located at the
anhydride groups of the PTCDA, right in the middle of the mentioned gap.
These orbitals fit into the level ordering scheme of DFTB, therefore it seems that LC-DFTB just shifts the levels of  DFTB if the range-dependent HF exchange term is added.
However, we find for other systems that the shift is non-uniform and the level ordering is not preserved 
(see below the discussion of pentacene). For long-range corrected
functionals  the transition from $\omega=0$ (DFT limit) to finite $\omega$ values 
is quite generally accompanied by smooth inhomogenous shifts of the single particle
levels. Therefore level reordering is expected. Obviously the
BNL/cc-pVTZ theory correctly exhibits this reordering of levels with respect to PBE/cc-pVTZ, 
which is not observed for LC-DFTB in the example shown here. This
should be attributed to the typical DFTB approximations, like the reduced
basis set and the two-center approximation, since already the DFTB
results in Figure \ref{fig:levels}d) show strongly underbound
$\sigma$-orbitals.  As level ordering issues have also been observed
in first principles LC-DFT, there can also be an additional effect.
The analysis in Ref.\ \citenum{Koerzdoerfer2012} showed, that even if the frontier orbitals seem to be well described by LC-DFT, there are 
states, usually of different symmetry (e.g. $\sigma$-orbitals), which exhibit considerable orbital SIE (OMSIE). 
In the aforementioned work the spectrum of tuned LC-$\omega$PBE functional has been discussed.
Within the tuning procedure the value of the range-separation parameter is chosen such 
that the HOMO eigenvalue is equal to the ionization energy 
obtained from the difference of total energies of the neutral species
and the cation. For this functional the second peak in the 
PTCDA spectrum is composed of degenerate $\sigma$ states (which
correspond to the HOMO-2/HOMO-3 in the LC-DFTB spectrum in the present work).
Analysis of the orbital self-interaction error for this theory showed large OMSIE in the $\sigma$-orbitals, 
while for $\pi$-orbitals the error was small.
LC-$\omega$PBE with standard value of the range-separation parameter 
exhibited the opposite behavior, where the $\sigma$-orbitals had rather small OMSIE. 
Thus the failure of LC-DFTB could also be connected to residual
self-interaction error which is more pronounced for the anhydride
$\sigma$-orbitals. In fact, the energetical position of the $\pi$
orbitals relative to the HOMO level is quite well represented by
LC-DFTB, much better than in first principles PBE.

The problematic level ordering of LC-DFTB is also observed for the pentacene spectrum (right part of Figure \ref{fig:levels}). 
While the levels up to HOMO-4 in DFTB, PBE and BNL show the same order, 
the LC-DFTB spectrum is characterized by the appearance of two $\sigma$-orbitals at HOMO-3 and HOMO-5 positions.
They are indicated by red and blue lines respectively.
It turns out that these orbitals  have already a differing position in the DFTB spectrum as compared to PBE. 
The approximate theories tend to underbind these orbitals.
Thus in this case the influence of the DFTB approximations is more evident. 
From the analysis above, we conclude
that the level ordering problem of the LC-DFTB might be partially caused by the  orbital self-interaction error of the $\sigma$-states
within the LC-DFTB theory as well as by the applied DFTB
approximations and the minimal basis set used. Again we find a
significant spreading of the quasi-particle spectrum in line with the
LC-DFT results and quite accurate level positions for the HOMO, HOMO-1
and HOMO-2 levels of $\pi$-nature. Notwithstanding, a full
characterization of an experimental photoemission spectrum seems to be
too ambitious at this point.

At the end of this section we briefly comment on the computational
efficiency of the new scheme versus first principles approaches. In fact,
the main motivation to use approximate methods like LC-DFTB is the
possibility to investigate large systems well beyond the scope of
conventional DFT codes.    
In Section \ref{impl} we already documented the quadratic scaling of
the method with system size. In Table \ref{tab:photovolttime} absolute timings of single point calculations for the theories considered in the previous discussion are summarized. 
The LC-DFTB calculations, with $\epsilon_{\text{threshold}}=10^{-16}$ (denoted by LC-DFTB$^1$) and 
$\epsilon_{\text{threshold}}=10^{-6}$ (denoted by LC-DFTB$^2$)
have been performed on a single core of an Intel Core-i7 CPU as before. 
The execution time was measured by the Linux {\em time} utility, where
the user time in seconds has been collected. For the DFT calculations the NWCHEM-6.3 code was used in serial and parallel versions. 
The serial version of NWCHEM-6.3 was executed on Intel Xeon 2.8GHz
machines, while the parallel jobs were distributed over 12 CPUs on a
cluster. Wall times were extracted from the NWCHEM-6.3 output files.

As expected, the first-principles calculations are computationally more demanding.
Even the calculation with small basis set at the BNL/3-21G theory level is at least 30 times slower than LC-DFTB$^1$ and 50 times slower than LC-DFTB$^2$ for smaller molecules. 
We note that the threshold parameter $\epsilon_{\text{threshold}}=10^{-6}$ gives the eigenvalues with MAE errors below $10^{-5}$ eV (see Figure \ref{fig:scaling}), 
thus this choice can be considered as accurate for practical calculations. 
For larger systems the gap in computational time between the LC-DFTB and first-principles calculations 
increases due to the quadratic scaling of the LC-DFTB method. 
At the same time traditional DFTB is at least an order of magnitude faster than LC-DFTB with $\epsilon_{\text{threshold}}=10^{-6}$. 

\begin{table*}
  \centering
\begin{tabular}{lrrrrrr}
\hline
\hline
 molecule     &  BNL/3-21G  &  BNL/cc-pVDZ  &  BNL/cc-pVTZ  &  LC-DFTB$^1$  &  LC-DFTB$^2$ & DFTB  \\
\hline
 5-Acene      &      325  &       2396  &      59701      &                 11  &      6  &  1   \\
 Perylene     &      541  &       4447  &     113414      &                 12  &      6  &  1  \\
 H$_2$P       &      932  &       7486  &     23825${^*}$  &                 16  &      8  &  1  \\
 Coronene     &     2116  &      15983  &     260677       &                 14  &    6   &  1  \\
 6-Acene      &      507  &       3516  &      79993       &                 15  &     8  &  2  \\
 5-Thiophene  &     1303  &      11581  &     144735       &                 16  &    11  &  2  \\
 PTCDA        &     3748  &      20477  &     521424       &                 22  &   10   &  2  \\
 H$_2$Pc      &     3034  &      27231  &    366838${^*}$  &                 39  &    14  &  2   \\
 H$_2$TPP     &     5735  &      43304  &    744967${^*}$  &                 72  &    22  &  4   \\
 C$_{60}$      &     7221  &      65393  &    655789${^*}$  &                121  &   23   &  2  \\
\hline
\hline
\end{tabular}
  \caption{Wall time [s] of DFT calculation versus LC-DFTB  for
 molecules with more than 30 atoms from the considered set.
The DFTB results are given for comparison as well.
The asterisk denotes the parallel jobs on 12 CPUs,
LC-DFTB$^1$ was performed with $\epsilon_{\text{threshold}}=10^{-16}$ and LC-DFTB$^2$ with $\epsilon_{\text{threshold}}=10^{-6}$. }
  \label{tab:photovolttime}
\end{table*}

\subsection{Electric field response: DFTB vs. LC-DFTB}
\label{response}
A well known deficiency of local DFT is the exaggerated response to an
applied electrostatic field. This behavior have been attributed to the
lack of a necessary non-local response term in the
exchange-correlation
functional\cite{Gonze1995,vanGisbergen1999,Karolewski2009}.  Thus all
local and semilocal exchange-correlation functionals fail to produce
the correct induced field, which is opposed to the applied electric field. 
As a consequence, local DFT leads to a wrong density distribution, 
characterized by a too strong separation of the induced charge.
This results in a drastic overestimation of static polarizabilities,
which gets stronger with growing system size. 
Difficulties amplify for the hyperpolarizability and second hyperpolarizability \cite{Champagne1998}. 
This problem has also consequences for the application of DFT in the field of molecular electronics. 
The underestimated HOMO-LUMO gap, lack of a field-counteracting term
and delocalization of the density lead altogether to a flawed 
description of transport properties, such as conductance.\cite{Lindsay2007}
We showed already in the section \ref{sec:quasipart}, that the LC-DFTB provides essentially better description of the 
fundamental gap, which suggests the reduction of the delocalization problem as compared to traditional DFTB. 
Recently, Sekino et al.\ provided evidence that LC-DFT shows the tendency to overcome the field response problem.\cite{Sekino2007}  
We therefore seek to confirm the signatures of a field-counteracting term due to
the inclusion of the non-local range-dependent term in the LC-DFTB method. 
We calculate the static longitudinal polarizabilities of polyacetylene
chains (PA, C$_{2n}$H$_{2n+2}$) with varying number of unit cells $n$ and analyze the induced Mulliken charge distribution along the chain. 
LC-DFTB as well as traditional DFTB include the electric field
$\mathbf{F}$ via the additional contribution
\begin{align}
  E_{\text{field}} = -\sum_A \Delta q_A \mathbf{F}\cdot\mathbf{R}_A
\end{align}
to the total energy functional, where $\Delta q_A = q_A - q^0_A$ is
the difference of the Mulliken population 
$ q_A = \sum_{\nu\in A}\left(\mathbf{P}\mathbf{S}\right)_{\nu\nu}$ and
the number of valence electrons on atom $A$ located at position $\mathbf{R}_A$. 
We point out that
traditional DFTB (and for the same reasons LC-DFTB) shows in general poor performance in predicting polarizabilities. 
This is due to the minimal basis set employed. However, LC-DFT even
with minimal basis (e.g. BNL/STO-3G) tends to correctly reduce the polarizability 
with respect to LDA/GGA-based theory in the same basis.

\begin{figure}
  \centering
  \includegraphics[width=0.8\textwidth]{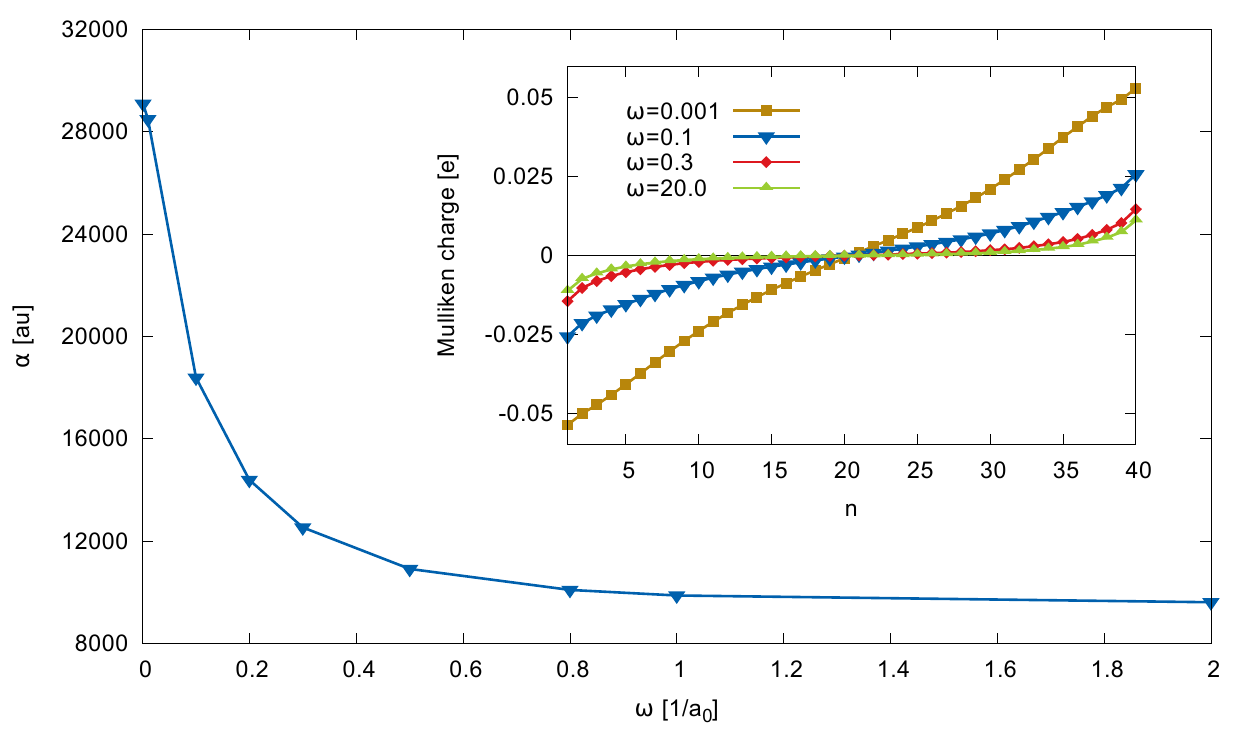}
  \caption{LC-DFTB linear polarizability of the PA ($n$ = 40) oligomer as a function of range-separation parameter $\omega$. 
The induced Mulliken charge per unit (F = 0.001 a.u.) for different values of
$\omega$ is shown in the
inset. 
}
  \label{fig:chgpol}
\end{figure}

We compare the polarizabilities obtained from the LC-DFTB method to long-range corrected DFT at the
BNL/6-311G** and BNL/3-21G level. For the case of LC-DFT the polarizabilities have been obtained from coupled-perturbed Kohn-Sham theory (CPKS), implemented in the NWCHEM-6.3 package.
The geometries for PA with $n$ = 10 and $n$ = 40 have been optimized at the B3LYP/6-311G* level of theory. 
We obtain the LC-DFTB polarizabilities by applying the finite field method. 
The numerical derivative of the dipole moment $\mu$ along the long
axis with respect to the perturbing electric field $F$ is calculated
with the center difference formula $\alpha = (\mu(F) - \mu(-F))/2F$,
where the field strength was chosen to be F = 0.0004 a.u.. 
\begin{table}[t]
  \centering
\begin{tabular}{lccc}
\hline
\hline
    $\omega [a^{-1}_0]$  &  BNL/6-311G**  &  BNL/3-21G  &  LC-DFTB  \\
\hline
   2.0  &          1288  &       1229  &     1156  \\
   0.8  &          1212  &       1138  &     1200  \\
   0.5  &          1193  &       1102  &     1256  \\
   0.3  &          1321  &       1215  &     1345  \\
   0.2  &          1513  &       1400  &     1429  \\
   0.1  &          1809  &       1696  &     1560  \\
  $10^{-2}$  &      2058  &       1938  &     1698  \\
  $10^{-3}$  &      2059  &       1939  &     1701  \\
\hline
\hline
\end{tabular}
  \caption{Static longitudinal polarizability of PA ($n$ =
    10). All values are in atomic units [$a^2_0 e^2/E_h$]. 
}
  \label{tab:10pa_pol}
\end{table}
Table \ref{tab:10pa_pol} lists the longitudinal static polarizability of
the PA oligomer with $n$ = 10 for BNL/6-311G**, BNL/3-21G and LC-DFTB at different values of the range-separation parameter $\omega$. 
We observe similar qualitative behavior, although the quantitative differences are rather large, especially for the local DFT limit ($\omega\to 0$). 
In this local DFT limit all three theories exhibit larger
polarizabilities as in the opposite HF+c limit ($\omega\to\infty$). In
the latter case the exchange-correlation functional is composed of 100 \% HF exchange and local DFT correlation. 
Remarkable is the rapid drop of the polarizability with the increase of the range-separation parameter $\omega$ (compare also Figure \ref{fig:chgpol}), 
seen in all theories. 
While in the case of $n$ = 10 the ratio of HF+c to local DFT limits
is 0.63 for both BNL/6-311G** and BNL/3-21G, and 0.68 for LC-DFTB, 
it decreases to 0.23 for BNL/3-21G and 0.33
in the case of LC-DFTB for the larger system with $n$ = 40 units. 
This indicates the aforementioned growing of the polarizability overestimation with the increasing system size. 

Further information is obtained by inspection of the charge
density. The inset of Figure \ref{fig:chgpol} depicts the induced
Mulliken charge due to a field of magnitude F = 0.001 a.u.
along the $n$ = 40 oligomer for different values of the range-separation parameter $\omega$.  
The LC-DFTB($\omega\to 0$) 
shows an almost linear charge distribution, which indicates an overly large polarization. 
We note that the DFTB is quantitatively very close to the LC-DFTB in the DFTB limit $\omega\to 0$, although the 
exchange-correlation functional is slightly different. Thus we do not show the DFTB result for brevity. 
Increase of the parameter $\omega$ gives rise to an effective screening of the electric field, 
which leads to the correction of the polarizability towards more physical values. 
The LC-DFTB polarizability as a function of the range-separation
parameter is provided in the main part of Figure \ref{fig:chgpol}
for comparison. We emphasize the qualitative nature of the LC-DFTB
results presented here. Quantitatively correct polarizabilities
require in general large basis sets. An alternative solution is the exploitation of empirical correction methods \cite{Kaminski2012,Kaminski2012b}.

\subsection{Proteins in gas phase and solution}
\label{proteins}
Experimental techniques that allow for the non-destructive extraction of proteins in combination with structural 
analysis make it possible to study intramolecular interactions in the absence of a particular solvent.\cite{Shelimov1997}  
Electrospray ionization (ESI)\cite{Fenn1989} together with mass
spectrometry and ion mobility measurements permit gas-phase structure
determination. Likewise, recent progress in the development of X-ray
free-electron lasers hold the promise to resolve protein structure in
vacuum at atomic resolution.\cite{Chapman2006,Meyer2013} From the
theoretical side, computations using solvent models in addition to gas
phase calculations may finally provide a way to understand the protein folding mechanism
in different environments. 

Efforts in this direction may address the still controversial question
whether peptides adopt the zwitterionic form known from aqueous
solution also in the gas
phase.\cite{Suenram1980,Price1997,Marchese2010}  DFTB, like other DFT
approaches based on local xc-functionals,  shows difficulties in the
description of the zwitterionic state where long-range charge-charge interactions play an important role. 
In a recent study Nishimoto et al.\cite{Nishimoto} found that the
DFTB self-consistency cycle failed to converge for the model peptides
chignolin (PDB ID: 1UAO\cite{Honda2004}) and Trp-cage (PDB ID: 1L2Y\cite{Neidigh2002}) in the
zwitterionic conformation. It has been, however, shown both
theoretically with force-field based molecular dynamics and
experimentally using electrospray ionization and photo-dissociation, 
that the native zwitterionic configuration of Trp-cage remains stable in the gas-phase \cite{Patriksson2007,Kjeldsen2006}. 

\begin{figure}
  \centering
  \includegraphics[width=0.8\textwidth]{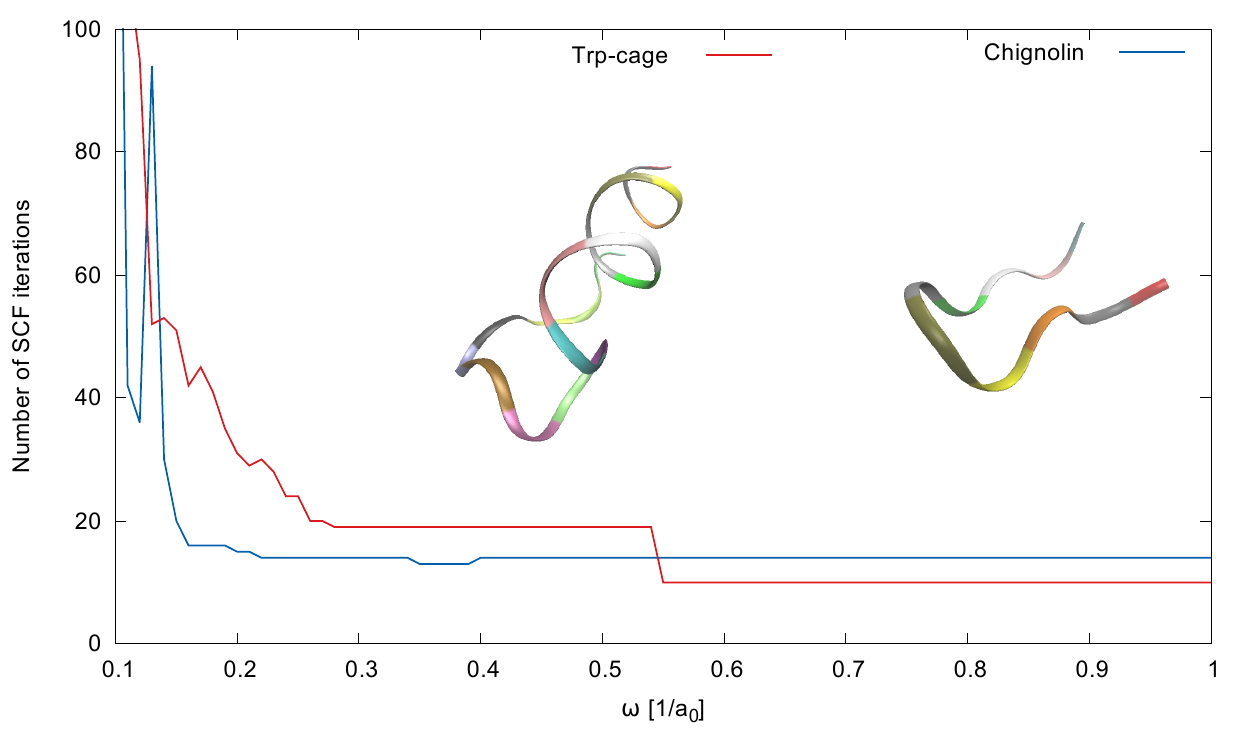}
  \caption{The SCF convergence plot of chignolin and Trp-cage proteins 
 as a function of range-separation parameter $\omega$ for the LC-DFTB
 method. The SCF does not converge in the local DFTB limit ($\omega=0$). 
Structures of the proteins are shown as insets. 
These plots have been generated from PDB
structures\cite{Neidigh2002,Honda2004} with the VMD software\cite{HUMP96}.
}
  \label{fig:prot1}
\end{figure}

Raising the electronic temperature seems
to circumvent the issue as a technical workaround, 
but in fact cures the symptoms instead of solving the fundamental
problem, which is the notorious underestimation of the HOMO-LUMO gap
of traditional DFTB. Motivated by this we present single-point calculations on the 1UAO and 1L2Y structures using the LC-DFTB method and compare to the 
LC-PBE/3-21G and RHF/3-21G methods in the following.  The geometrical
structure is in both cases the native zwitterionic
conformation, obtained by NMR
measurements as deposited in the protein data base.\cite{Neidigh2002,Honda2004} For our gas phase
simulations we consider the models to be charge neutral in total. To
this end basic and acidic side chains have been restored to their
neutral form by appropriate protonation or deprotonation.\footnote{The geometry
of the affected functional groups was subsequently optimized at the
DFTB level (electronic temperature T = 500 K) leaving the rest of the
molecule fixed.}  The carboxyl
and amino terminals were kept oppositely charged. A ribbon representation of both proteins is given in Figure \ref{fig:prot1}.

The LC-DFT and RHF calculations have been performed with the parallel version of NWCHEM-6.3 on 8 CPUs, whereas 
LC-DFTB was run on a single core.
As can be seen from the Table \ref{tab:hlgprotein} a good agreement 
in the description of the frontier orbital energies 
between full DFT and LC-DFTB is achieved. The timings can be found in
the bottom part of the Table. We also performed LC-DFTB calculations
for different values of the range-separation parameter $\omega$. The
convergence (number of SCF cycles) for both proteins is depicted in
the figure \ref{fig:prot1}
. As can be clearly seen, the convergence
issue does not occur for
typical values of the range-separation parameter. As the band gap opens, convergence generally improves.

\begin{table}[t]
  \centering
  \begin{tabular}{lrrrrrr}
\hline
\hline
         & \multicolumn{3}{c}{Chignolin} & \multicolumn{3}{c}{Trp-cage} \\
         & HOMO & LUMO & Gap & HOMO & LUMO & Gap \\
\hline
LC-DFTB($\omega=0.3$)                 &  -3.15 & 0.63 & 3.78 & -3.66  & -1.18  & 2.48 \\
LC-DFTB($\omega\rightarrow\infty$)    &  -5.90 & 2.14 & 8.04 & -6.14  &  1.32  & 7.46 \\
LC-PBE/3-21G                          &  -3.78 & 0.34 & 4.12 & -4.46  & -1.72  & 2.74 \\
RHF/3-21G                             &  -5.42 & 1.81 & 7.23 & -6.12  & -0.13  & 5.99 \\
\hline
\hline
         & \multicolumn{3}{c}{time [sec]} & \multicolumn{3}{c}{time [sec]}  \\
\hline
LC-DFTB($\omega=0.3$)  & \multicolumn{3}{c}{  73} & \multicolumn{3}{c}{ 882} \\
LC-DFTB($\omega\rightarrow\infty$) & \multicolumn{3}{c}{  78} & \multicolumn{3}{c}{  331} \\
LC-PBE/3-21G           & \multicolumn{3}{c}{ 8841} & \multicolumn{3}{c}{36844} \\
   RHF/3-21G           & \multicolumn{3}{c}{ 1577} & \multicolumn{3}{c}{ 7534} \\
\hline
\hline
  \end{tabular}
  \caption{Frontier orbital energies and fundamental gap (all in eV)
    of the chignolin and Trp-cage zwitterions in the gas-phase for
    different theories. 
}
  \label{tab:hlgprotein}
\end{table}

These results indicate that LC-DFTB allows for reliable studies of
proteins in the gas-phase both in the neutral and zwitterionic
conformation. Certainly, further studies need to address also the
energetics of the various conformers but already at this point LC-DFTB
seems to be a useful tool for biological applications. In this context
it might serve as the underlying electronic structure method for the
fragment molecular orbital approach (FMO)
\cite{Fedorov2012,Nagata2011,Nishimoto2014}, which enables the
quantum chemical study of biological systems with many thousands of
atoms.  
\section{Conclusion and outlook}
In the present paper we addressed the implementation and benchmark of a long-range corrected exchange-correlation 
functional in the DFTB method, which we denote as LC-DFTB, for
closed-shell systems.  The practical implementation requires the
extension of tools for the evaluation of Slater-Koster parameters as
well as some rather minor changes to the DFTB Hamiltonian. 
Thus the scheme can be easily implemented in existing {\it ab initio} software packages, 
which in principle contain all necessary routines for the integral evaluation and Hamiltonian construction.

The Hamiltonian construction exhibits quadratic scaling and dominates 
the computational time for the systems tested. 
For larger systems the method is expected to show cubic scaling due to the 
diagonalization. The performance benefit with respect to first
principles DFT with long-range corrected functionals still remains two to
three orders of magnitude depending on the size of the basis set.

LC-DFTB shows clear signatures of correction of the delocalization
problem in local DFT that is attributed to the self-interaction error. 
This has been demonstrated for the frontier orbitals of a set of
organic molecules, where in general a promising agreement with 
full LC-DFT with double zeta basis (BNL/3-21G) has been
observed. Remaining flaws were related to the minimal basis set
characteristic for DFTB, which influences especially electron affinity
levels. As a second example we calculated the static longitudinal
polarizability of polyacetylene chains and provided evidence for the 
qualitatively correct description of the response potential. 

The parametrization of the repulsive potential, which requires an
adjustment due to the modified Hamiltonian, is presently on the
way. With this development, thermochemical and structural properties
may be investigated in addition to the electronic structure. The
extension of the method to a spin-unrestricted formalism is likewise
promising, since it would allow for the non-empirical tuning of the
range-separation parameter.\cite{Baer2010} Finally we would like to
point out that the presented formalism is not restricted to the
specific long-range corrected functional used in this study. The
scheme can be easily adapted to conventional hybrid functionals like
B3LYP or to screened exchange functionals like HSE,\cite{Heyd2003} which are recently
becoming increasingly popular for periodic systems and solid state
applications.     
 
\section{Acknowledgement}
Financial support from the Deutsche Forschungsgemeinschaft (GRK 1570) is gratefully acknowledged. We thank Stephan Irle and Yoshio Nishimoto (Nagoya University) 
for making the authors aware of the potential applications of the
presented method in the field of biological systems.
\appendix
\section{Note on the numerical evaluation of integrals}
\label{app:integrals}
The evaluation of electron repulsion integrals for Slater type
orbitals (STO) as used in the DFTB method is a numerically challenging
task. For the Coulomb interaction several algorithms based on the analytical expansions of STOs around a center are available.
Recently, Seth and Ziegler \cite{Seth2012} extended
these techniques to the case of a Yukawa interaction ($\exp(-\omega
r)/r$). In the present method the run-time evaluation of integrals is
explicitly avoided by invoking the Slater-Koster scheme with
precomputed parameters. Thus the integral evaluation does not affect
the computational performance. We therefore resort to a simple, yet
robust, numerical integration scheme proposed by Becke \cite{becke:2993}.\footnote{The choice of the numerical
  integrator can be additionally motivated by the fact that the DFTB
  method does in principle not rely on 
a specific type of orbitals as basis functions. 
This allows for more flexibility, especially in the context of atomic
DFT calculations with converged numerical orbitals.}   
This scheme was designed for the evaluation of two-electron integrals
over numerically defined charge distributions with Coulomb interaction. 
The modification for the case of Yukawa interaction is straightforward
and laid out below.
 
The double integral over four arbitrary orbitals $\phi_i(\mathbf{r})$, which are in general located at different centers, and some interaction $g(|\mathbf{r}-\mathbf{r}^\prime|)$
\begin{align}
  (ab|cd) = \int\int \phi_a(\mathbf{r})\phi_b(\mathbf{r})\,g(|\mathbf{r}-\mathbf{r}^\prime|)\phi_c(\mathbf{r}^\prime)\phi_d(\mathbf{r}^\prime)\; d\mathbf{r}d\mathbf{r}^\prime,
\end{align}
can be seen as a special case of the integral over general charge distributions $\rho(\mathbf{r})$, $\sigma(\mathbf{r})$ 
\begin{align}
\label{eq:generalintegralbecke}
  I = \int\int \rho(\mathbf{r}) \sigma(\mathbf{r}^\prime) g(|\mathbf{r}-\mathbf{r}^\prime|) \; d\mathbf{r}d\mathbf{r}^\prime
    = \int \rho(\mathbf{r}) \underbrace{ \left[\int
        \sigma(\mathbf{r}^\prime) g(|\mathbf{r}-\mathbf{r}^\prime|)
        d\mathbf{r}^\prime \right] }_{V(\mathbf{r})} d\mathbf{r}.
\end{align}
The potential $V(\mathbf{r})$ can be obtained by solving the following
boundary value problem with $V(\mathbf{r}) \rightarrow 0, |\mathbf{r}|\rightarrow \infty$ 
\begin{align}
  \mathcal{D} V(\mathbf{r}) &= -4\pi \sigma(\mathbf{r}) 
\end{align}
with some differential operator $\mathcal{D}$.  
The function 
$g(|\mathbf{r}-\mathbf{r}^\prime|)$ in eq. \ref{eq:generalintegralbecke} is  the Greens function of the problem
\begin{align}
  \mathcal{D} g(|\mathbf{r}-\mathbf{r}^\prime|) &= -4\pi \delta(\mathbf{r}-\mathbf{r}^\prime). 
\end{align}
The differential operator for the Coulomb interaction $g(r) = 1/r$ 
is $\mathcal{D} = \nabla^2$. 
The Yukawa interaction $g(r) = \exp(-\omega r) / r$ 
is the Greens function for the modified Helmholtz operator $\mathcal{D} = \nabla^2 - \omega^2$. 
Thus the Poisson equation in the original work by Becke has to be replaced by the modified Helmholtz equation. 

To make use of efficient quadrature schemes, Becke proposed a space partitioning and decomposition of the charge distribution into  
atom-centered portions. The space is partitioned by some set of functions $f_A(\mathbf{r})$, 
with $\sum_A f_A(\mathbf{r}) = 1,\;\forall \mathbf{r}\in\mathbb{R}^3$. 
 Using this set of functions the density is divided into the atomic contributions
  \begin{align}
    \rho(\mathbf{r}) = \sum_A f_A(\mathbf{r}) \rho(\mathbf{r}) = \sum_A \rho_A(\mathbf{r}).
  \end{align}
 This allows to write the integral in Eq.\
 \ref{eq:generalintegralbecke} as a sum of two-center integrals over atom-centered charge distributions $\rho_A,\sigma_B$
 \begin{align}
   I = \sum_{AB} I_{AB} =  \sum_{AB} \int\int \rho_A(\mathbf{r}) \sigma_B(\mathbf{r}^\prime) g(|\mathbf{r}-\mathbf{r}^\prime|)\; d\mathbf{r}d\mathbf{r}^\prime.
 \end{align}
The inner part of the two-center integral $I_{AB}$ is evaluated as follows.
In the Helmholtz equation 
\begin{align}
\label{eq:helmholtzbecke}
 \left( \nabla^2 - \omega^2 \right) V_A(\mathbf{r}) = -4\pi \rho_A(\mathbf{r}),
\end{align}
the potential and density are expanded into spherical harmonics 
\begin{align}
\label{eq:potexp}
  V_A(\mathbf{r}) &= \sum_{lm} r^{-1} V_{A,lm}(r) Y_{lm}(\Omega) \\
  \rho_A(\mathbf{r}) &= \sum_{lm} \rho_{A,lm}(r) Y_{lm}(\Omega).
\end{align}
Inserting these expansions into Eq.\ \ref{eq:helmholtzbecke} gives a set of one-dimensional radial equations 
\begin{align}
 \left[ \frac{d^2}{dr^2} -\frac{l(l+1)}{r^2} - \omega^2\right] V_{A,lm}(r) = -4\pi r \rho_{A,lm}(r),
\end{align}
with boundary conditions $\lim_{r\to 0}V_{A,lm}(r) = 0,\; \lim_{r\to \infty}V_{A,lm}(r) = 0$. 
These equations are solved using a finite difference method and from the resulting $V_{A,lm}$ the 
potential $V_A(\mathbf{r})$ is assembled according to  Eq.\ \ref{eq:potexp}. 
The remainder of the integration is performed on a two-center grid as
described in the original paper by Becke.\cite{becke:2993}

\section{Analytical evaluation of the long-range $\gamma$-integral}
\label{sec:apphubbard}
The $\gamma$-integral over the Yukawa interaction can be reduced to a one-dimensional integration\cite{Niehaus2012}
\begin{align}
  \gamma^{\text{Y},\omega}_{AB}(R_{AB}) &= \frac{\tau^3_A \tau^3_B}{(8\pi)^2}
\int\int e^{-\tau_A |\mathbf{r}_{1}-\mathbf{R}_A|}\frac{e^{-\omega |\mathbf{r}_{1} - \mathbf{r}_{2}|}}{|\mathbf{r}_{1} - \mathbf{r}_{2}|} e^{-\tau_B |\mathbf{r}_{2}-\mathbf{R}_B|}\;d\mathbf{r}_1 d\mathbf{r}_2  \\
\label{eq:1dlrgamma}
 &=\frac{2\tau^4_A \tau^4_B}{\pi R_{AB}} \int^\infty_0 \frac{q \sin(q
   R_{AB})}{(q^2+\tau^2_A)^2(q^2+\tau^2_B)^2(q^2+\omega^2)} dq,
\end{align}
where $R_{AB} = |\mathbf{R}_A - \mathbf{R}_B|$.
This integral can now be further evaluated using the residue theorem. 
We obtain then the analytical expression\footnote{
The analytical formula Eq.\ \ref{eq:anlrgamma} has been derived after the calculations for this work have been done.  
For the presented results the long-range integral has been evaluated with a numerical integrator.
We note that the analytical formula is more practicable and is recommended for use in practical implementation of the method.
}
\begin{align}
\label{eq:anlrgamma}
  \gamma^{\text{Y},\omega}_{AB}(R_{AB}) 
&= \frac{\tau^4_A \tau^4_B}{(\tau^2_A - \omega^2)^2(\tau^2_B - \omega^2)^2}\frac{e^{-\omega R_{AB}}}{R_{AB}} \nonumber \\
&-\Bigg[e^{-\tau_A R_{AB}}\left(\frac{\tau^2_A}{\tau^2_A - \omega^2}\frac{\tau_A \tau^4_B}{2(\tau^2_B - \tau^2_A)^2} 
- \frac{\tau^4_A}{(\omega^2 - \tau^2_A)^2}\frac{(\tau^6_B + 3\tau^2_A \tau^4_B + 2\omega^2 \tau^4_B)}{(\tau^2_A - \tau^2_B)^3 R_{AB}}\right) \nonumber \\
& + e^{-\tau_B R_{AB}}\left(\frac{\tau^2_B}{\tau^2_B - \omega^2}\frac{\tau_B \tau^4_A}{2(\tau^2_A - \tau^2_B)^2} 
- \frac{\tau^4_B}{(\omega^2 - \tau^2_B)^2}\frac{(\tau^6_A + 3\tau^2_B \tau^4_A + 2\omega^2 \tau^4_A)}{(\tau^2_B - \tau^2_A)^3 R_{AB}}\right) \Bigg].
\end{align}
In the limit $\omega\to 0$ the integral is taken over the Coulomb interaction. 
In this case our analytical formula reduces to the result of Elstner et al.\ \cite{Elstner1998}. 
The long-range $\gamma$-integral is then given 
by the difference $\gamma^{\text{lr}}_{AB} = \gamma^{\text{Y},0}_{AB} - \gamma^{\text{Y},\omega}_{AB}$. 

To obtain the 
on-site value ($R_{AB}\to 0,\;\;A=B$), we go back to the one-dimensional integral Eq.\ \ref{eq:1dlrgamma}. 
We expand the term $\sin(qR_{AA})$ around zero and integrate again using the residue theorem
\begin{align}
 &\lim_{R_{AA}\to 0} \frac{2\tau^8_A }{\pi R_{AA}} \int^\infty_0 \frac{q \sin(q R_{AA})}{(q^2+\tau^2_A)^4(q^2+\omega^2)} dq \\
\label{eq:singint}
 &= \frac{2\tau^8_A}{\pi}\int^\infty_0\frac{q^2}{(q^2+\tau^2_A)^4(q^2+\omega^2)} dq \\
 &= \frac{\tau^8_A}{48(\tau^2_A - \omega^2)^4}\left[\frac{3(5\tau^6_A + 15\tau^4_A \omega^2 - 5\tau^2_A \omega^4 + \omega^6)}{\tau^5_A}-48\omega \right].
\end{align}
This yields the on-site value for the integral Eq.\ \ref{eq:gammalr}
\begin{align}
\label{eq:gammalr_onsite}
  \gamma^{\text{lr}}_{AA} = \frac{5}{16}\tau_A - \frac{\tau^8_A}{(\tau^2_A - \omega^2)^4}\left[\frac{5\tau^6_A + 15\tau^4_A \omega^2 - 5\tau^2_A\omega^4 + \omega^6}{16\tau^5_A} -\omega\right].
\end{align}

\section{The Hubbard parameter in LC-DFTB}
\label{sec:hubpar}
We derive the expression for the Hubbard parameter of atomic LC-DFTB. 
Note that in this case $S_{\mu\nu} = \delta_{\mu\nu}$, $\gamma^{\text{fr}/\text{lr}}_{\mu\nu} = \gamma^{\text{fr}/\text{lr}}_{AA}$ and $H^0_{\mu\nu} = \delta_{\mu\nu} \epsilon_\mu$. 
Using the orhogonality condition $\sum_\mu c_{\mu, i}c_{\mu, j} = \delta_{ij}$ we obtain the total LC-DFTB energy of one atom in terms of occupation numbers $n_i$
\begin{align}
  E^{\text{atom}} &= \sum_{\mu} P_{\mu\mu} \epsilon_\mu + \frac{1}{2}\gamma^{\text{fr}}_{AA} \sum_{\mu\alpha}\Delta P_{\mu\mu} \Delta P_{\alpha\alpha} 
 - \frac{1}{4}\gamma^{\text{lr}}_{AA} \sum_{\mu\nu}\Delta P_{\mu\nu} \Delta P_{\mu\nu}  \\
&= \frac{1}{2}\sum_{ij} n_i n_j  \gamma^{\text{fr}}_{AA} - \frac{1}{4}\sum_{i} n^2_i   \gamma^{\text{lr}}_{AA} + \text{terms linear in }n_i 
\end{align}
We consider further only terms, quadratic in the occupation numbers, which belong to the highest occupied shell. 
It contains $n$ electrons, equally distributed over the shell. 
The occupation number for an orbital is then $n/d_l$, where $d_l=2l+1$ is the degeneracy of the shell with angular momentum $l$. 
Then the energy can be written as 
\begin{align}
E^{\text{atom}} &=\frac{1}{2}\gamma^{\text{fr}}_{AA}\sum\limits^{\text{shell}}_{ij}\left(\frac{n}{d_l}\right)^2 - \frac{1}{4}\gamma^{\text{lr}}_{AA} \sum\limits^{\text{shell}}_{i} \left(\frac{n}{d_l}\right)^2 
+ \text{terms linear in }n \\
& =\frac{1}{2}\gamma^{\text{fr}}_{AA} d^2_l \left(\frac{n}{d_l}\right)^2 - \frac{1}{4}\gamma^{\text{lr}}_{AA} d_l \left(\frac{n}{d_l}\right)^2 + \text{terms linear in }n \\
&=\frac{1}{2}\gamma^{\text{fr}}_{AA} n^2 - \frac{1}{4} \gamma^{\text{lr}}_{AA} \frac{n^2}{d_l} + \text{terms linear in }n
\end{align}
Thus the second derivative with respect to the occupation numbers of the highest occupied shell is 
\begin{align}
  \frac{\partial^2 E^{\text{atom}}}{\partial n^2} = \gamma^{\text{fr}}_{AA} - \frac{1}{2}\frac{1}{2l+1}\gamma^{\text{lr}}_{AA}
\end{align}
And together with Eq. \ref{eq:gammalr_onsite} the atomic Hubbard from the LC-DFTB is obtained
\begin{align}
  U_A = \frac{5}{16}\tau\left[1-\frac{1}{2(2l+1)}\left(1-\frac{\tau^8 + 3\tau^6\omega^2 - \tau^4\omega^4 + 0.2 \omega^6\tau^2 - 3.2\tau^7\omega}{(\tau^2-\omega^2)^4}\right)\right]
\end{align}
This equation defines the decay constant $\tau$. 

\bibliographystyle{ieeetr} 
\bibliography{combined}

\begin{figure}[H]
 \centering
  \includegraphics[width=\textwidth]{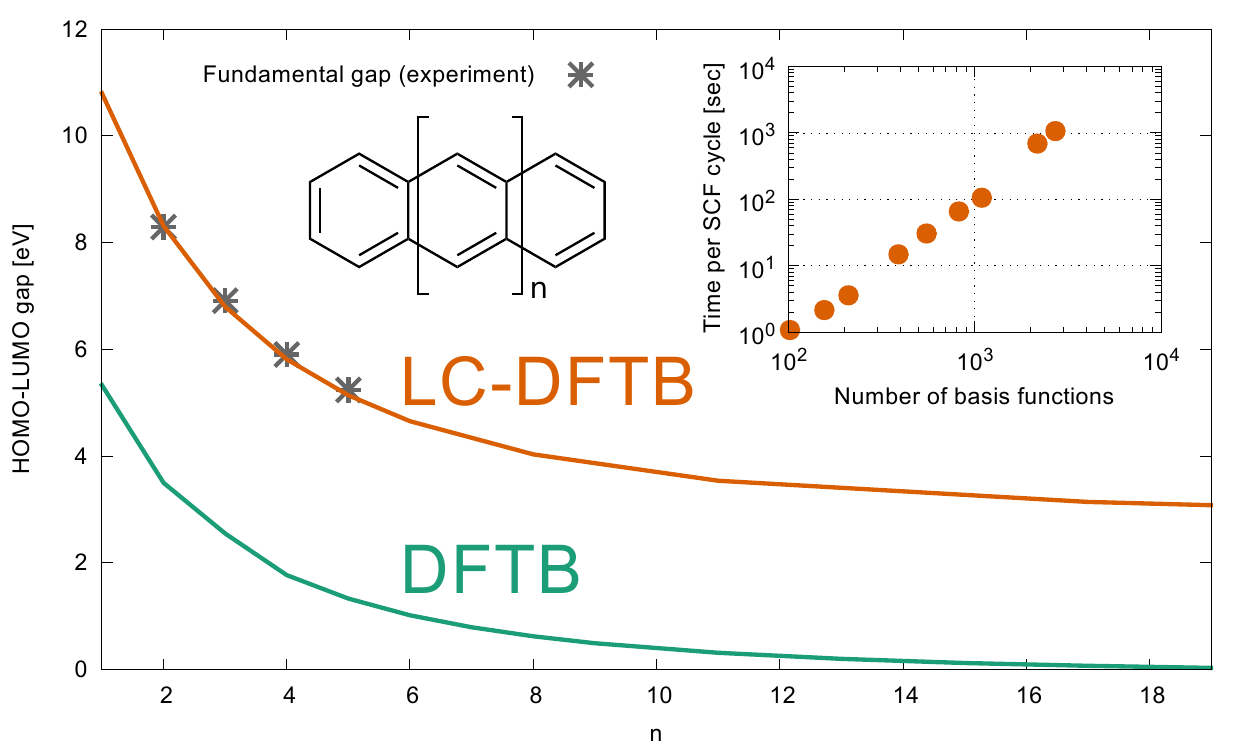}\\
  \caption{Table of Contents Graphic} 
  \label{toc}
\end{figure}

\end{document}